# Criteria and Approaches for Virtualization on Modern FPGAs

Duc-Canh Le, Chan-Hyun Youn
*Department of Electrical Engineering*
*Korea Advanced Institute of Science and Technology, Daejeon, Korea*
{canhld, chyoun}@kaist.ac.kr

*Abstract*—**Modern field programmable gate arrays (FPGAs) can produce high performance in a wide range of applications, and their computational capacity is becoming abundant in personal computers. Regardless of this fact, FPGA virtualization is an emerging research field. Nowadays, challenges of the research area come from not only technical difficulties but also from the ambiguous standards of virtualization. In this paper, we introduce novel criteria of FPGA virtualization and discuss several approaches to accomplish those criteria. In addition, we present and describe in detail the specific FPGA virtualization architecture that we developed on Intel Arria 10 FPGA. We evaluate our solution with a combination of applications and microbenchmarks. The result shows that our virtualization solution can provide a full abstraction of FPGA device in both user and developer perspective while maintaining a reasonable performance compared to native FPGA.**

## I. INTRODUCTION

In computing domain, hardware virtualization refers to creating multiple logical hardware components from a physical one. Virtualization enables sharing one resource among many users, therefore, it helps increasing resource utilization, reducing cost and enabling the flexibility of the system. Over the past two decades, hardware virtualization has become increasingly popular as a core technology in many commercial solutions and products, for instance, virtual machines (VMs) and cloud computing. Nowadays, advances on hardware microarchitecture from chip vendors enables virtualizing CPU and memory with extremely low overhead [1]. On the other hand, I/O virtualization is dealing with significant challenges due to the heterogeneous characteristic of I/O devices. Field programmable gate arrays (FPGAs) in particular is an example of an I/O device with a combination of high efficiency, structure complexity, irregular execution model and limited document[1].

Going away from being chips in embedded systems or application specific integrated circuits (ASICs) prototyping, FPGAs have become a general-purpose platform for serving different types of workload including machine learning, image processing and scientific computing in recent years. Taking the advantages of small size, low power, and runtime reconfigurable, enterprises start adding FPGAs as a general component in their datacenter for accelerating either I/O connections [2][3] or business services [4]. Meanwhile, FPGAs computational performance is rapidly growth, outperforms CPU and comes closer to graphical co-processors (GPUs) performance. Modern SRAM-based FPGAs provide ever richer programmable logic resources of miscellaneous types [5] and being able to host multiple accelerators in a single chip. On the other hand, many challenges in hardware development on FPGA have been primarily addressed. High-level synthesis enables programming on FPGAs with high level programming languages such as C, C++ or Java. FPGA OpenCL software development kit provides an abstraction on FPGA structure and brings FPGA closer to the community. Of all this broad scope of developments makes it progressively essential to disclose FPGAs to virtualized environment.

Unfortunately, virtualization of FPGAs is extremely complicated. There are distinct natures of FPGAs that prevent emulating it by a software model, which we categorize as *heterogeneity*, *irregularity*, and *infidelity*. First of all, the heterogeneity is about the mixture of resources. In micro-structure, a FPGA is an array of thousands of computational resources, connected by a programmable routing bus system. Those resources include look-up tables, registers, digital signal processors, memory blocks and hard peripheral controllers. Because of the density, either emulating the resources or the connectivity on the software is unrealistic. Secondly, the irregularity is on the execution model. Most of general-purpose processors do the computation in time domain. Program instructions are step by step fetched from memory and issued to a single or a group of an execution unit. In contrast, FPGAs unroll all instructions, creates a hardware primitive for each instruction and executes the program on space domain. Using a temporal model to emulate FPGAs certainly results in terrible performance. Finally, the infidelity is due to current capabilities of FPGA chips. Hardware on FPGA is created by defining the resources and programing the connection among them. The description of the hardware is stored in a bitfile which can be pre-configured or loaded at run-time. Within current development tool chains and FPGA technologies, a program is synthesized to different bitfiles on different FPGAs or even different regions on an FPGA.

Because of these natures, techniques and approaches using in CPU, memory, or common I/O virtualization are not directly applicable. Nevertheless, FPGA virtualization has been studied since the 1990s in the signs of dynamic loading [6], partitioning, and overlay [7]. The concept of virtualization has been changed over time due to the rapid change of application requirements [8]. Hence, the definition and criteria of FPGA virtualization are diverse and ambiguous. Non-standardization has negative impacts on the field of research, i.e. difficulties for new researchers or misleading among virtualization and other technology.

Obviously, it difficult to have a complete standard of FPGA virtualization at this moment. However, motived by previous works on CPU and GPU [9][10], and in an attempt to shed some light on it, this paper present novel criteria and taxonomy of strategies for FPGA virtualization. We also describe and evaluate our virtualization architecture, which is developed on Intel FPGA, as a case study for our proposed goals. In summary, we make the following contributions:

- We proposed novel criteria of FPGA virtualization and taxonomy of virtualization approaches.
- We provided a case study on FPGA virtualization following our criteria on current FPGA.
- We analyzed the cost of virtualization on modern FPGA from both hardware and software perspective.

---

[1] There are different strategies of FPGAs deployment in modern computing system, however, in this paper we consider coupling FPGAs with a CPU, which is the most general case.

The remaining of the paper is organized as follows. Section II briefly provide some backgrounds and summaries related works. Section III describes our virtualization criteria and discusses approaches to accomplish those criteria. Section IV describes the design and evaluation of our virtualization architecture, Section V concludes the paper.

## II. RELATED WORKS

FPGA virtualization has been studied since the 1990s however, the motivation and approaches are changed through time. In the early time, there are works to support *multitasking* on FPGAs. Specifically, Dehon et al. introduced the concept of dynamically programmable gate arrays (DPGAs) which allowed runtime reconfiguration on FPGAs [6]. The authors presented several useful design patents on FPGA and showed how they can bring area efficiency gains on DPGA. Simmler et al. proposed an FPGA management unit and a client-server model to enabling multitasking on general computers with FPGA co-processor [11]. Christian et al. suggested *virtualized execution* model with PipeRench [12]. The author assumed a system with $n$ reconfigurable processors and $m$ applications where $m > n$. PipeRench decomposes programs to sequential steps, and both the processing and the reconfiguration are pipelined to provide the illusion of smooth execution to each application. Multitasking is simple but also costly in time due to the reconfiguration overhead. In addition, the system must be preemptive for efficient multitasking, which is difficult on FPGAs. Though there are recent works to reduce the cost of hardware checkpoint [13][14], FPGA virtualization on time domain is, in general, a drawback because it omits the spatial execution model of FPGA [15].

Within the growing of Linux systems, there are efforts to develop an *operating system for reconfigurable hardware* by abstracting FPGAs in Linux kernel. The abstract layer handles the reconfiguration of FPGAs, the hardware schedule and the communication between hardware and software. Therefore, FPGAs are neglected from user's view. Notable works include BORPH, OS4RS, and ReconOS [16]–[18]. On the other hand, *overlay* accomplishes the abstraction by introducing an intermediate fabric on the top of an FPGA [19]–[22]. Overlay provides an extra level of programmability and productivity but comes at a substantial cost of routing congestion [8].

*Virtual FPGA* is an emerging concept coming in the cloud era. Taking the advantages of dynamic partial reconfiguration, a modern FPGA is divided into smaller regions that can be reconfigured at run-time while the rest of the device is still running. By adding hardware *shell* that provides the interface to the outside world, each region appears to users as a physical FPGA. Most remarkable work from IBM Research developed a virtualized FPGA system and integrated it to Linux-KVM and OpenStack to bring virtual FPGA to the cloud [5]. There are similar efforts from the University of Warwick, TU Dresden, EPFL [23]–[25] and University of Manchester [15].

In general, virtual FPGA comes with some costs that need to be justified. Firstly, the routing congestion is not as serious as that on the overlay, however, it is difficult to achieve an economical timing design. S. Yazdanshenas and V. Betz show in [26] that network on chip (NOC) is a promising solution. Secondly, the number of virtual FPGA and the floorplanning of the design is essential to achieve performance and equality among users. There is an effort from [27] to provide automatic and feasible floorplanning in partial reconfiguration system. Finally, the ecosystem on virtual FPGA, including libraries, toolchains and development kits, is starving. Without the ecosystem, developers and users obviously do not want to build their products on virtual FPGA.

## III. FPGA VIRTUALIZATION CRITERIA AND APPROACHES

### A. Virtualization goals

Motivated by Popek's goals of x86 machine virtualization [9] and Dowty's goals of GPU virtualization [10], we propose five criteria for judging FPGA virtualization: *performance*, *fidelity*, *multiplexing*, *isolation,* and *interposition*.

The motivation of *performance* and *fidelity*, which are taken from [9], is minimizing the cost of virtualization. In general, performance denotes the relative speed of programs, and fidelity means the preservation of features on virtualized devices, compared to the native one. Specifically, for the performance, we consider three facts: (1) FPGAs have spatial execution model, therefore a resource overused program cannot fit on FPGAs; (2) virtualized FPGA may have less resource than physical FPGA in many strategies; and (3) even a same program is synthesized to different hardware between virtualized and physical FPGA. Hence, our criterion implies that within a design which can be synthesized on both a virtualized and a native FPGA, the virtualized FPGA should run at a comparable speed to that on the native FPGA. Higher performance improves the user's experience in virtualized environments. Meanwhile, *fidelity* suggests that advanced features on physical FPGA, e.g., reconfigurability and high-speed I/O connectivity, and the ecosystem should be retained in the virtualized environment. The ecosystem includes toolchains, libraries, software development kit (SDK) and even the design flow on FPGA. In other words, we believe that good virtualization strategies provide a complete abstraction of FPGAs not only in the user's but also in the developer's perspective.

*Multiplexing* underlines the extra value of virtualization. Multiplexing allows sharing a physical FPGA among threads, so a single FPGA can host multiple VMS. Within the growth of cloud computing in recent years, multiplexing becomes particularly essential in emerging business models, including *infrastructure as a service* (IaaS) or *accelerator as a service* (AaaS).

Finally, *isolation* and *interposition* are required to bring virtualized FPGA to the commercial environment. Isolation is the concept in machine virtualization as one VM cannot access others memory space. Isolation is an essential feature for enabling security in the public domain. Nevertheless, isolation in FPGA virtualization is extremely challenging because FPGAs are designed for a single user and they do not have a memory management unit. Though there are techniques to ease the impact and enable FPGAs in the kernel, isolation is best addressed by manufacturers through additional hardware change [5]. Interposition is the ability of recording accesses between the VMs and physical device with software. High level of interposition empowers advanced features in modern cloud data-center, for instance, VM live migration, checkpoint and restore. Furthermore, it is important to note that the concept of interposition does not include the hardware state in FPGAs within current technology.

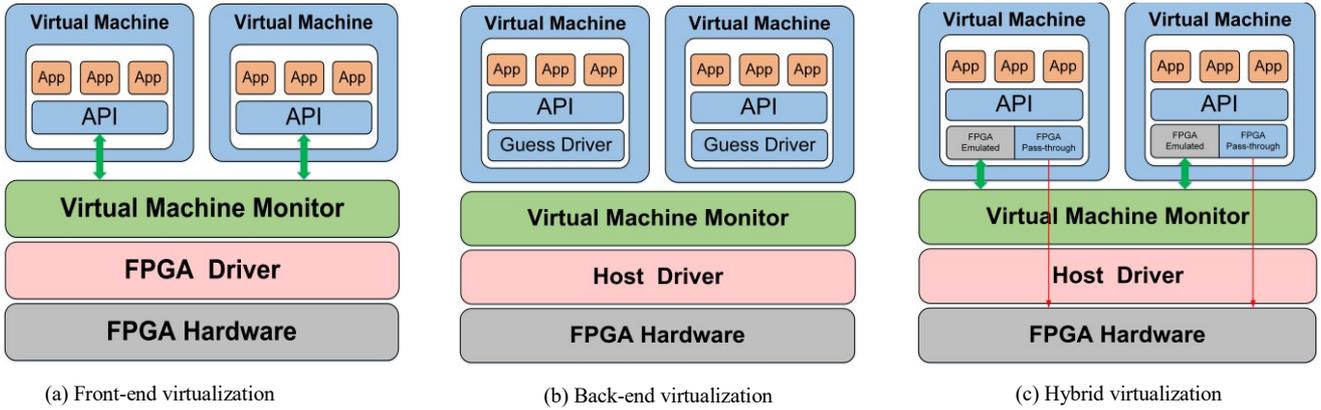

Figure 1: Virtualization strategies from high-level view

Next, we present the taxonomy of virtualization strategies. There is the tradeoff among these criteria, e.g. fidelity and performance are in opposition. Moreover, applications have different requirements, so the weigh on the criteria is different. Therefore, rather than ranking the criteria, we highlight the strengths and weakness of each strategy and specify suitable applications. Following [10], we generalize virtualization techniques into two approaches: front-end virtualization and back-end virtualization.

### B. Front-end virtualization

The front-end virtualization (FEV) in FPGAs shares some ideas with x86 para-virtualization, where the virtualization boundary is at fairly high-level of I/O stacks. Figure 1a shows a common architecture of FEV. Application program interface (API) is wrapped in high-level libraries and run in the guess, while the actual implementation is in the hypervisor. During the system runtime, requests from libraries are intercepted by the guess and redirected to virtual machine monitor (VMM). VMM receives requests from VMs and issues these requests to FPGA by an appropriate scheduling algorithm. Hence, the VMM plays the role of a resource broker between VMs and the physical FPGA.

Since there is no hardware and device driver modification, this approach is not specific to an FPGA model or vendor. It was shown in [28] that FEV can deliver good performance when all VMs use the same accelerator and no reconfiguration on FPGAs is needed. In addition, accesses to FPGA are totally mediated by the hypervisors, so the isolation among VMs is guaranteed and the interposition is done easily if desired. However, FEV also produces some major drawbacks. If the VMs are desired to use different accelerators, they need to reconfigure the FPGA. Meanwhile, the reconfiguration time of modern FPGAs could be very large, e.g. about 2.5 seconds over PCIe on Intel Arria 10 FPGA. Within this overhead, VMM is not able to schedule the requests. Consequently, FEV sacrifices the reconfigurability in exchange for performance. The other minus of FEV is the guess operating system itself should be modified to provide the hyper calls to the VMM. Therefore, FEV is suitable when FPGA is deployed in a system, not for general purposes, e.g. compression engine or packets co-processors.

### C. Back-end virtualization

The back-end virtualization (BEV) creates a virtualization boundary at low-level of I/O stack, generally between the driver and the physical FPGA. In Figure 1b, VMs have their own guess driver and may interact directly with the hardware. Therefore, this strategy possibly brings high performance and a fairly high level of fidelity to virtual FPGA. The earliest implementation of BEV is fixed pass-through where an FPGA is permanently attached to a VM. This solution has been exploited by cloud vendors for years to offer FPGA instances to the clients. Although fixed pass provides perfect isolation, nearly native performance and VMs can fully utilize the capabilities of the hardware, it has potential limitations on multiplexing and interposition. In addition, coupling a physical FPGA with a VM is visibly expensive and not a general solution.

As we mentioned before, FPGAs themselves do not support multi-context execution. However, taking advantage of partial reconfiguration, modern FPGAs can be partitioned into independence partitions called *partial reconfiguration regions* (PRRs). By adding a shell to provide interfaces to these partitions, each of them can host an accelerator, hence enables concurrency on FPGAs. Shell actually is the area outside of the PRRs (or *static region*), which implement the communication infrastructure to connect PRRs to the outside world. Partial reconfiguration can be exploited in BEV to provide multi-tasking by dedicating just a PRR rather than an entire FPGA to VM and emulating the common components such as DMA and reconfiguration controllers. This solution shares similar themes with mediated pass-through in GPU virtualization [10].

Although mediated pass-through solves the multiplexing, it suffers additional cost: the host/hypervisor driver should be customized to efficiently manage the PRRs and common resources. In addition, the logical FPGA which appear in the VM may have different hardware interfaces compared to the original FPGA. That means modifications may be required on the guess driver, and the ecosystem on physical FPGA can be lost. Furthermore, hardware synthesized on PRR-based virtualized FPGA is obviously not optimized due to many physical constraints, that may influence the performance. FPGAs also do not implement memory management mechanism, so each PRR can potentially interference others and break down the isolation.

In sum, BEV takes a further step toward reconfigurable hardware as a service on virtualized machines. However, there are many concerns on the isolation or security, the fidelity, and the performance. Therefore, our virtualization as shown in Figure 1c combines FEV and BEV to gain the benefits of both approaches. By combining both BEV and FEV, it is possible to use FEV for security-sensitive operations, e.g. FPGA memory access, reconfiguration, and DMA, and pass-through is utilized to provide access to each PRRs from VMs. The detail of structure and implementation are introduced in the following Section.

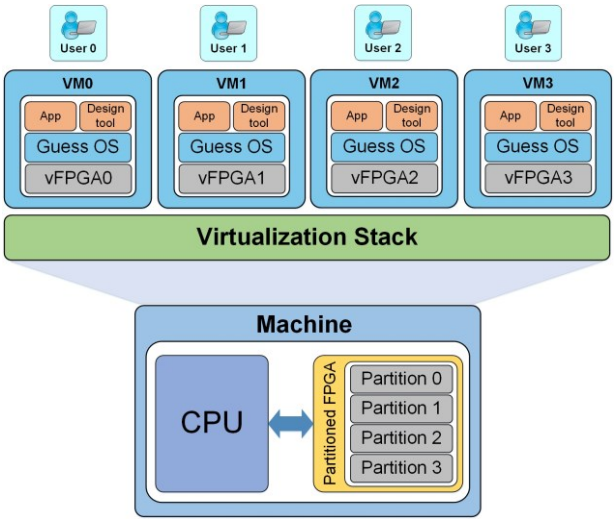

Figure 2: Application scenario of virtualization in the paper

## IV. FPGA VIRTUALIZATION PLATFORM

### A. Application scenario

There are many scenarios that FPGA virtualization can bring benefits. In this paper, we consider a use case of sharing an FPGA among users in a local cluster as shown in Figure 2. The user of each VM owns a virtual FPGA (vFPGA), actually is a part of the physical FPGA, and the virtualization stack should give the illusion to users that they are using a physical FPGA. Specifically, our platform was designed within the following bindings in mind. First of all, virtualization must deliver full features of physical FPGA to vFPGAs. That means vFPGAs are reconfigurable and users can create their own accelerators on vFPGAs with the same design flow that being used on the native FPGA. Users also are able to use the same programming interface to develop applications and run their accelerators on vFPGAs. In other words, design toolchain and runtime libraries should be preserved in the VM environment. Next, virtualization must provide reasonable performance on vFPGAs, compared to native FPGA. The performance is evaluated by both application benchmarks and micro-benchmarks. The application benchmarks demonstrate the overall speed of vFPGA, while the micro-benchmarks stress different components of the system: the PCIe bandwidth, the vFPGA memory bandwidth, and the frequency of vFPGA. Finally, virtualization must produce isolation among vFPGAs, e.g. in Figure 2 user of VM0 cannot access to the memory region used by vFPGA1.

Since our virtualization solution provides users the illusion of physical FPGA on a vFPGA, either education or business can benefit from it. The most obvious application is in cloud computing, where providers can build FPGA instances by associating a VM with a vFPGA to offer users more options than *fixed pass-through*. The solution can also be applied to share the development kit among developers in the early stage of a product or students in a class to decrease the infrastructure cost.

### B. Hardware Layer

Our platform is extended from Intel reference design for OpenCL on Arria 10 development kit [29]. The original hardware from Intel comes with partial reconfiguration enabled and one PRR. We expanded the number of PRRs by re-floorplanning the design and adding the interfaces to the PRRs. The final design of our hardware is shown in Figure 3a. Especially, the key components that we customized are the IRQ handler, the PRR controller, the PRR interface and the clock generator.

The PRR controller is responsible for reconfiguring the PRRs. The core of this component is an FPGA control block (CB), which performs most of the important features: bitfile decoding, CRC checking, and PR flow handling. In addition, we add registers to provide a frozen interface to each PRR. The freeze signal is asserted at the beginning of the reconfiguration process to reset the PRR internal state and freeze all of the interfaces to this PRR and is deserted when the reconfiguration is done. The freeze signals are essential to avoid any harming signal from going to other modules when PRR is configuring.

We implement interrupt for synchronizing the operation between PRRs and the host CPU. The PCIe endpoint in Arria 10 device supports message signaled interrupt (MSI), and we use one MSI line for all PRRs. Especially, the IRQ controller concatenates the interrupts from PRRs, buffers them in a register, and generates the MSI signal. When the host receives the MSI, it will read the status register in the IRQ controller to detect the interrupt source and run the corresponding interrupt service routine (ISR). The IRQ controller also implements a control register to mask the interrupt when the host runs the ISR or when some PRRs are inactive.

In the Intel FPGA OpenCL platform, the PRR interface provides a control domain to the accelerator in the PRR. The accelerators (or the kernels) in the Intel FPGA OpenCL platform are generated from OpenCL toolchain and hence, have an identical register set. The PRR interface implements an intermediate buffer between the host and the PRR registers. In our platform, we want to keep an identical design flow to users on vFPGA. Therefore, we use the OpenCL compilation flow to generate the HDL code and write our own scripts to integrate the HDL code to each PRRs, synthesize the design and assemble the bitfile. To support multiple PRRs, we duplicate the PRRs interface in the shell instead of customizing the component. The underlying reason is for lowering the routing efforts. A PRR interface has direct connections to a PRR, so they should be placed close to each other. The floorplanning of other components is also carefully handled to avoid routing congestion. For example, the DDR bridges are placed close to the DDR pin bank; and the pipeline stages are added between the memory interfaces of PRRs to the DDR bridges to reduce the length of the critical path. In general, proper floorplanning can greatly reduce the routing length in our design, as shown in Figure 3b. However, the root of problems comes from the wide data buses in virtualization flow, as well as the spanning of connections from PRRs to the resources in the shell. Recent works present the potential of using the hard network on chip (NoC) as the interconnect solution for virtualized FPGAs to reduce the cost of routing.

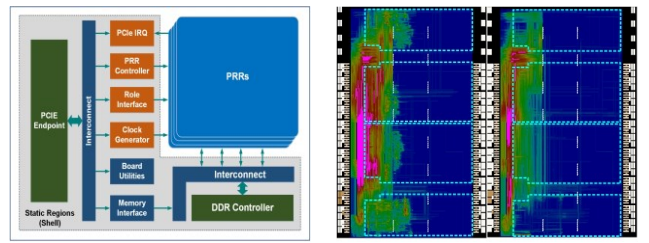

(a) Block design      (b) Comparison of floorplanning

Figure 3: Hardware design in virtualization platform

Finally, the clock generator is modified to generate four clock signals to PRRs. However, our customized clock generators cause compilation crash when running assembler, and it is believed to be related to a Quartus bug in version 17.1. Therefore, we temporary use a single clock to drive all of PRRs with the frequency of 200MHz.

*C. Software Stack*

We preserve the software stack of OpenCL platform on the virtualized environment with three layers: device driver, memory mapped device (MMD) library and OpenCL runtime library. Particularly, the device driver creates the device file in the user space and implements the file operators: *open*, *close*, *read* and *write*. The MMD library divorces the device file into several independent interfaces, and also implement the basic interface operators: *open*, *close*, *read*, *write*, *get_info*, *set_irq*, *set_status*, and *reprogram*. Here, *get_info* gets the information of the interface in the hardware, *set_irq* sets the interrupt handler that the MMD layer calls when receive interrupt signal from the driver, and the *set_status* set the status handler that is called when a special event finishes (e.g. at the end of data transfer from host to FPGA). Finally, the OpenCL runtime library implements the OpenCL API that is friendly to the users. Obviously, these layers should be modified to be usable on the VMs. The runtime library is closed source, however, Intel claims that the runtime layer works with any MMD layer which properly implements the basic interface operators.

Figure 4 shows the overall of our virtualization stack. On a VM, only one associated PRR to that VM appears to the user as a vFPGA and any access to this vFPGA is simply passed-thought. The rest of MMD APIs is forwarded to the VMM including the reprogram function and all of the memory related operations. The underlying reason for this design is to prevent the interferences among vFPGAs. Specifically, giving the reprogram permission to VMs can be dangerous because a malicious user can easily reprogram the vFPGA of other VMs. In PR-based virtualization, the bitfile that configured to vFPGA actually is a partial bitfile in PR compilation flow. The FPGA PR control block cannot check whether a partial bitfile is associated with a particular PRR but only the compatible to the device and the shell. Therefore, if a user in VM0 calls reprograming but uses the bitfile compiled for PRR1, the vFPGA in VM1 is reconfigured. In the same manner, a PRR has full access to the DRAM. Therefore, if the memory operations are implemented in VM, a user can, accidentally or intentionally, read others data.

In our virtualization stack, the reprogram requests in VM, the pointer to the desired bitfile, and the size of bitfile is forward to VMMs. The VMM will check the information embedded in the bitfile to confirm whether the requests are legal or not. The information of PRRs can be embedded in the bitfile easily in the complication process (that is hidden to users). On the other hands, a virtual memory management unit (MMU) is implemented in the VMM to handle the FPGA memory allocation requests (i.e. *clCreateBuffer* in OpenCL context). The idea of software MMU is same as memory virtualization in the CPU. We divide the memory of the FPGA board into numerous 1MB segments. When the allocation request comes, the MMU finds in the segments pool the first group of contiguous segments which satisfies the amount of the request and return the address of the first portion in this group. The implementation of the pool is simply as an array with free segments marked 0 and used segments marked 1. The algorithm can be further improved by using a linked list. For the data transferring, we use the VM-copy mechanism, mean the data is first copied from VMs memory to host memory, then moved to FPGA memory using DMA. In the future, VM-nocopy mechanism can be used to reduce the copy overhead.

The software workaround can only protect the user's data from the software side. If a user is malicious, he can create a hardware module to access a specific memory are. Therefore, even user's data cannot be read by others, it is potentially able to be damaged. The problem can be temporarily fixed by adding hardware modules to restrict the available memory area of each PRR. However, this solution will reduce memory utilization and in general, not a sufficient method. Overall, we believe that protecting user's data on the hardware side with low cost is hard and could be an interesting research topic.

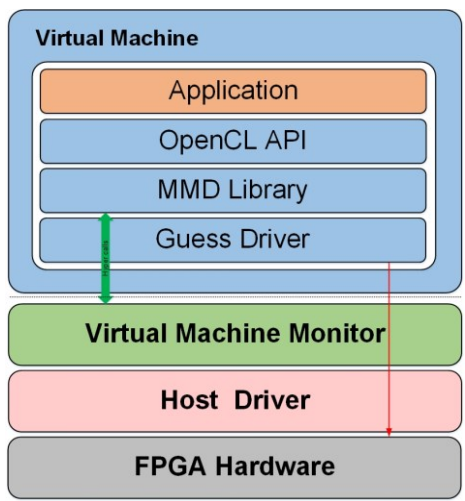

Figure 4: The software stack of virtualization

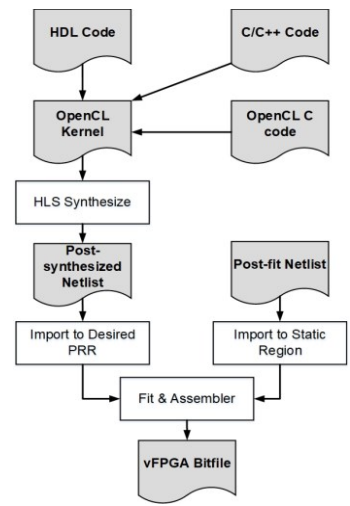

Figure 5: The compilation flow on vFPGAs

## D. Design flow on vFPGAs

As shown in Figure 5, the complication flow on vFPGA is actually the PR complication flow. However, we want to hide the partial reconfiguration and provide an identical design tool to the user. In this work, we modified the Intel's scripts to import the accelerator design to the desired PRR. The Intel OpenCL compilation program calls these scripts to create hardware, therefore it is available on the VMs.

## E. Evaluation

The evaluation is conducted on our server with an Intel Xeon E5-2609. The processor has 8 cores, operates at 1.7GHz. The server has total 32GB of DDR3 memory and runs Ubuntu 14.04. The FPGA kit using in our experiment is the Arria 10 development kit. We evaluate our virtualization with three applications: matrix multiplication, Sobel filter, and vector addition. In each evaluation, we generate the hardware from same OpenCL code and compared the performance between the vFPGA and the physical FPGA. We also perform micro-benchmark to identify the main bottle-neck of the platform. However, since adding customized software modules to exist VMM is fairly challenge and time consuming. Therefore, the evaluation in the paper is primarily performed on an emulation environment of our architecture, where the applications run in foreground processes, while MMU and the reconfiguration module run in a background process.

Figure 6a shows the running time of three programs on native FPGA and vFPGA. Overall, performance of vFPGA is lower than native FPGA in all applications. This degradation may come from various source: the frequency of accelerators, the data transfer time, or the software overhead. We break down the performance of vFPGA with vector addition in Figure 6b. The result shows that software computation time accounts for approximately 55% of the total running time. This is understandable because our virtualization platform introduces lots of software components to fix the hardware errata. Hence, more software optimization should be done in the future to improve the efficiency of the virtualization platform.

## V. CONCLUSION

In this paper, we analyze the characteristics of FPGA as an I/O device and propose novel criteria of FPGA virtualization. We further discuss and present a taxonomy the virtualization approaches in previous works. We provide a prototype of a virtualization platform that can bring FPGA to the virtualized environment with full abstraction in both user and developer perspective. We also demonstrate how our proposed criteria can be achieved in using our virtualization platform. Our future works will focus on the optimization of the software, the protection of user data on the hardware side and also how to bring our design into an existing VMM.

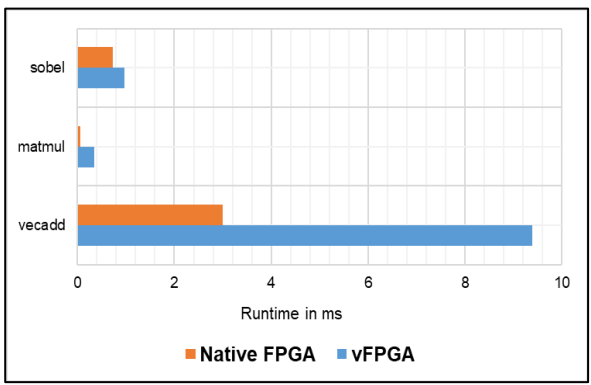

(a) Performance of applications on native FPGA and vFPGA

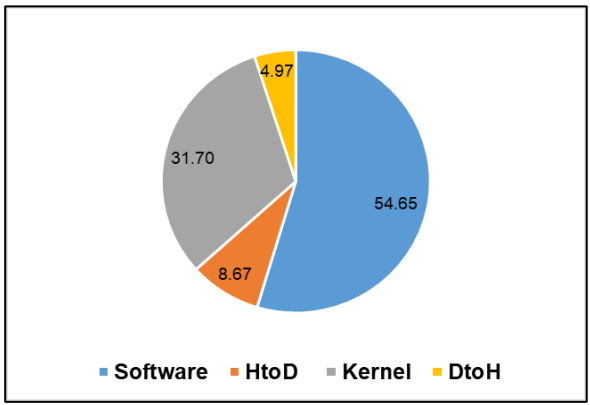

(b) Break down of performance on vFPGA

Figure 6: The evaluation result of virtualization platform

REFERENCES

[1] K. Adams *et al.*, "A comparison of software and hardware techniques for x86 virtualization," *ACM SIGOPS Oper. Syst. Rev.*, vol. 40, no. 5, p. 2, Oct. 2006.

[2] D. Firestone *et al.*, "Azure Accelerated Networking: SmartNICs in the Public Cloud," in *Nsdi'18*, 2018.

[3] Nallatech, "40Gbit AES Encryption Using OpenCL and FPGAs." pp. 3–6, 2016.

[4] E. Chung *et al.*, "Serving DNNs in Real Time at Datacenter Scale with Project Brainwave," *IEEE Micro*, vol. 38, no. 2, pp. 8–20, Mar. 2018.

[5] F. Chen *et al.*, "Enabling FPGAs in the cloud," in *Proceedings of the 11th ACM Conference on Computing Frontiers - CF '14*, 2014, pp. 1–10.

[6] A. DeHon, "DPGA Utilization and Application," in *Fourth International ACM Symposium on Field-Programmable Gate Arrays*, 1996, pp. 115–121.

[7] W. Fornaciari and V. Piuri, "General methodologies to virtualize FPGAs in Hw/Sw systems," in *1998 Midwest Symposium on Circuits and Systems (Cat. No. 98CB36268)*, 1998, pp. 90–93.

[8] A. Vaishnav, K. D. Pham, and D. Koch, "A Survey on FPGA Virtualization," in *2018 28th International Conference on Field Programmable Logic and Applications (FPL)*, 2018, pp. 131–1317.

[9] G. J. Popek and R. P. Goldberg, "Formal requirements for virtualizable third generation architectures," *Commun. ACM*, vol. 17, no. 7, pp. 412–421, Jul. 1974.

[10] M. Dowty and J. Sugerman, "GPU virtualization on VMware's hosted I/O architecture," *ACM SIGOPS Oper. Syst. Rev.*, vol. 43, no. 3, p. 73, Jul. 2009.

[11] H. Simmler, L. Levinson, and R. Männer, "Multitasking on FPGA Coprocessors," Springer, Berlin, Heidelberg, 2000, pp. 121–130.

[12] S. C. Goldstein, H. Schmit, M. Budiu, S. Cadambi, M. Moe, and R. R. Taylor, "PipeRench: a reconfigurable architecture and compiler," *Computer (Long. Beach. Calif).*, vol. 33, no. 4, pp. 70–77, Apr. 2000.

[13] D. Koch, C. Haubelt, and J. Teich, "Efficient hardware checkpointing," in *Proceedings of the 2007 ACM/SIGDA 15th international symposium on Field programmable gate arrays - FPGA '07*, 2007, p. 188.

[14] K. Rupnow, W. Fu, and K. Compton, "Block, Drop or Roll(back): Alternative Preemption Methods for RH Multi-Tasking," in *2009 17th IEEE Symposium on Field*

*Programmable Custom Computing Machines*, 2009, pp. 63–70.
[15] A. Vaishnav, K. D. Pham, D. Koch, and J. Garside, "Resource Elastic Virtualization for FPGAs Using OpenCL," in *2018 28th International Conference on Field Programmable Logic and Applications (FPL)*, 2018, pp. 111–1117.
[16] Hayden Kwok-Hay So, "BORPH: An Operating System for FPGA-Based Reconfigurable Computers," 2007.
[17] Chun-Hsian Huang and Pao-Ann Hsiung, "Hardware Resource Virtualization for Dynamically Partially Reconfigurable Systems," *IEEE Embed. Syst. Lett.*, vol. 1, no. 1, pp. 19–23, May 2009.
[18] E. Lübbers and M. Platzner, "ReconOS," *ACM Trans. Embed. Comput. Syst.*, vol. 9, no. 1, pp. 1–33, Oct. 2009.
[19] J. Coole and G. Stitt, "Intermediate fabrics," in *Proceedings of the eighth IEEE/ACM/IFIP international conference on Hardware/software codesign and system synthesis - CODES/ISSS '10*, 2010, p. 13.
[20] A. Brant and G. G. F. Lemieux, "ZUMA: An Open FPGA Overlay Architecture," in *2012 IEEE 20th International Symposium on Field-Programmable Custom Computing Machines*, 2012, pp. 93–96.
[21] D. Koch, C. Beckhoff, and G. G. F. Lemieux, "An efficient FPGA overlay for portable custom instruction set extensions," in *2013 23rd International Conference on Field programmable Logic and Applications*, 2013, pp. 1–8.
[22] C. Liu, H.-C. Ng, and H. K.-H. So, "QuickDough: A rapid FPGA loop accelerator design framework using soft CGRA overlay," in *2015 International Conference on Field Programmable Technology (FPT)*, 2015, pp. 56–63.
[23] S. A. Fahmy, K. Vipin, and S. Shreejith, "Virtualized FPGA Accelerators for Efficient Cloud Computing," in *2015 IEEE 7th International Conference on Cloud Computing Technology and Science (CloudCom)*, 2015, pp. 430–435.
[24] O. Knodel, P. Lehmann, and R. G. Spallek, "RC3E: Reconfigurable Accelerators in Data Centres and Their Provision by Adapted Service Models," in *2016 IEEE 9th International Conference on Cloud Computing (CLOUD)*, 2016, pp. 19–26.
[25] M. Asiatici, N. George, K. Vipin, S. A. Fahmy, and P. Ienne, "Virtualized Execution Runtime for FPGA Accelerators in the Cloud," *IEEE Access*, vol. 5, pp. 1900–1910, 2017.
[26] S. Yazdanshenas and V. Betz, "Quantifying and mitigating the costs of FPGA virtualization," in *2017 27th International Conference on Field Programmable Logic and Applications (FPL)*, 2017, pp. 1–7.
[27] M. Rabozzi, G. C. Durelli, A. Miele, J. Lillis, and M. D. Santambrogio, "Floorplanning Automation for Partial-Reconfigurable FPGAs via Feasible Placements Generation," *IEEE Trans. Very Large Scale Integr. Syst.*, vol. 25, no. 1, pp. 151–164, Jan. 2017.
[28] Wei Wang, M. Bolic, and J. Parri, "pvFPGA: Accessing an FPGA-based hardware accelerator in a paravirtualized environment," in *2013 International Conference on Hardware/Software Codesign and System Synthesis (CODES+ISSS)*, 2013, pp. 1–9.
[29] Intel, "Intel FPGA SDK for OpenCL - Overview." 2018.